\documentclass[letterpaper,twocolumn,10pt]{article}
\usepackage{usenix_style}

\usepackage{graphicx}
\usepackage{subcaption}
\usepackage{textcomp}
\usepackage{xcolor}
\usepackage{array}
\usepackage{tabularx}
\usepackage{hyperref}
\usepackage{tikz}
\usepackage{xspace} 
\usepackage{enumitem}
\usepackage{caption}
\usepackage{float}
\usepackage{booktabs}          
\usepackage{nicefrac}       
\usepackage{silence}
\usepackage{microtype}       
\usepackage{listings}
\usepackage{makecell}
\usepackage{amsmath,amssymb,amsfonts}
\usepackage{algorithm}
\usepackage{algorithmicx}
\usepackage{algpseudocode}
\renewcommand{\algorithmicrequire}{\textbf{Input:}}
\renewcommand{\algorithmicensure}{\textbf{Output:}}
\newcommand{\customTableFont}{\fontsize{7pt}{8pt}\selectfont}

\newenvironment{conditions}
  {\par\vspace{\abovedisplayskip}\noindent
   \tabularx{\columnwidth}{>{$}l<{$} @{${}:{}$} >{\raggedright\arraybackslash}X}}
  {\endtabularx\par\vspace{\belowdisplayskip}}
\begin{document}
%-------------------------------------------------------------------------------

\date{}

\title{\bf Efficient Resource Scheduling for Distributed
Infrastructures Using Negotiation Capabilities}

\author{
Junjie Chu\ \ \
Prashant Singh\ \ \
Salman Toor
\\
\\
\textit{Department of Information Technology, Uppsala University}
}

\maketitle

%-------------------------------------------------------------------------------
\begin{abstract}
In the past few decades, the rapid development of information and internet technologies has spawned massive amounts of data and information. 
The information explosion drives many enterprises or individuals to seek to rent cloud computing infrastructure to put their applications in the cloud. 
However, the agreements reached between cloud computing providers and clients are often not efficient. 
Many factors affect the efficiency, such as the idleness of the providers' cloud computing infrastructure, and the additional cost to the clients. One possible solution is to introduce a comprehensive, bargaining game (a type of negotiation), and schedule resources according to the negotiation results. 
We propose an agent-based auto-negotiation system for resource scheduling based on fuzzy logic. The proposed method can complete a one-to-one auto-negotiation process and generate optimal offers for the provider and client. 
We compare the impact of different member functions, fuzzy rule sets, and negotiation scenario cases on the offers to optimize the system. 
It can be concluded that our proposed method can utilize resources more efficiently and is interpretable, highly flexible, and customizable. 
We successfully train machine learning models to replace the fuzzy negotiation system to improve processing speed. 
The article also highlights possible future improvements to the proposed system and machine learning models. 
All the codes and data are available in the open-source \href{https://github.com/Junjie-Chu/Efficient_Resource_Scheduling_for_Distributed_Infrastructures_using_Negotiation_Capabilities}{repository}.
\end{abstract}
%-------------------------------------------------------------------------------

\section{Introduction}
In the past few years, the rapid development of information and internet technology has spawned massive amounts of data and information. 
It is not economical for most enterprises and individuals to update and maintain hardware devices for data-intensive applications. 
Therefore, many enterprises or individuals seek to rent cloud computing infrastructure to put their applications in the cloud to reduce costs and gain better flexibility. 
Many cloud computing infrastructure providers build lots of data centers and lease-related resources to meet the needs of the above tenants \cite{velte2010cloud,varghese2018next}. 
The most famous case of tenants is Netflix. The organization spent years migrating data from its data centers to AWS \cite{NetflixAWS}. 
Well-known representatives of infrastructure providers include Amazon Web Services, Microsoft Azure, and Huawei Cloud \cite{shimba2010cloud,roumani2019empirical}.

Cloud services have a range of advantages, but they are still far from perfect. 
For example, in the process of buying and selling, such a basic assumption is followed: the more consumers buy, the more discounts the sellers will give, and the lower the unit price of goods purchased by consumers. 
When infrastructure tenants and providers negotiate offers, tenants may underestimate the potential future resource demand and purchase resources at a higher unit price when further capacity expansion occurs. 
For the providers, the excess resources have been idle, and cannot make a profit.
In addition, idle resources create inefficient utilization of the resources which leads to unnecessary costs and carbon emissions \cite{jones2018stop, googlecloudcarbon}.

One possible new solution is to introduce a comprehensive, win-win autonomous bargaining process (a type of negotiation) and schedule resources according to the results of the process. 
In this comprehensive negotiation, the interests of both parties are resolved through a negotiation process oriented towards cooperation and information exchange. 
This is a well-known way to find win-win agreements between providers and tenants \cite{lobel1994realities}. On the condition that the basic requirements of the tenants are met, the participants regard the negotiation as a ``problem to be solved,'' then approach a solution acceptable to both parties through negotiation, and finally reach a win-win agreement. 
In the case of cloud computing resources, it may even be a win-win-win. In addition to tenants and infrastructure providers, the environment can benefit from more efficient resource utilization due to the negotiation game.

The solution suffers from the following two challenges:

\begin{itemize}
\item How to implement an effective agent-based autonomous negotiation system?   
\item How to speed up the entire negotiation process while keeping the offering mechanism explainable?
\end{itemize}

In this article, a negotiation system based on fuzzy logic is designed and developed. 
In this system, we design two agents to simulate different actors, roles, and responsibilities within distributed infrastructure settings. We have identified two main actors: infrastructure providers and application owners (tenants). 
The aim is to offer cost-effective execution plans to application owners while maximizing infrastructure providers' resource utilization and economic benefits. 
Storage, VCPU, and RAM will be considered parameters in the negotiation process. Multiple levels of information volumes are exchanged between the agents running on behalf of the tenants and suppliers. 
In the end, an offer that satisfies both parties based on a specific level of mutual information will be reached through negotiation. 
Moreover, multiple machine learning models are selected and trained, which can replace the negotiation game and achieve similar results with higher efficiency. 
It is important to note that the choice of these particular parameters is based on the way current service providers offer services. 
The presented framework is not limited to only these parameters.

Fuzzy negotiation systems are a promising solution to infrastructure resource scheduling. 
The negotiation function can be a dynamic entity better represented using fuzzy logic compared to crisp functions. Machine learning and fuzzy logic are considered two separate approaches to solving problems. They can combine but rarely replace each other. 
Therefore, another highlight of this article is the systematic presentation of the feasibility (probably the only one so far, as far as we know) of replacing fuzzy negotiation systems with machine learning.

In typical business activities, participants need trustworthy ways to make decisions. 
A fuzzy logic negotiation system could introduce important interpretability, which satisfies the requirements of business activities. 
Machine learning models viewed as black boxes, although difficult to interpret, provide very high efficiency. Our article has the best of both worlds and therefore has strong practicality in business.

The main contributions of the article are:

\begin{itemize}
\item A comprehensive fuzzy logic-based negotiation system that supports multiple cases and custom parameters to get offers and provide resource scheduling indications. 
\item A workflow that uses neural network models (surrogate model) to replace fuzzy negotiation systems to speed up.
\item Extensive experiments to prove the feasibility of the neural network model to replace the fuzzy negotiation system and compare the performance.
\item Overall, we propose a trustworthy, interpretable, and efficient scheduling method for resources that combines machine learning models and fuzzy negotiation.
\end{itemize}

The remainder of this article is organized as follows: \autoref{sec.preliminaries} presents the scenarios, concepts, and technical tools in the article. 
\autoref{sec.systemdescription} introduces the design of fuzzy negotiation systems and machine learning workflow. 
\autoref{sec.exp} explains the details of related experiments, including the settings and results.
\autoref{Related work} reviews related literature.
In \autoref{sec.conclusion}, we conclude the work, and further possible improvements are discussed.

%-------------------------------------------------------------------------------
\section{Preliminaries}
\label{sec.preliminaries}
%-------------------------------------------------------------------------------

%-------------------------------------------------------------------------------
\subsection{Scenario}
\label{Scenario}
%-------------------------------------------------------------------------------
Tenants who order large quantities tend to enjoy discounts on unit prices, which is a widely used promotion method in business activities. 
There is also such a situation in the field of infrastructure services. 
Whether it is Google Cloud or Huawei Cloud, they all provide more favorable unit prices to customers who rent a lot according to their price tables \cite{Googlecloud}\cite{Huaweicloud}. In our article, we imitate Huawei Cloud's pricing system and set up a simplified simulated one as \autoref{tab:price}.

In this article, we use \emph{client's agent} to represent tenants and \emph{provider's agent} to represent infrastructure service providers in our negotiation systems. 
The scenario can be described in detail as follows: 
After reading \autoref{tab:price}, the tenant inputs its requirements and other related parameters to the client's agent, and then the agent submits this information to the provider's agent directly. 
The provider's agent then negotiates with the client's agent to provide new offers based on the input provided by the client's agent. Compared with the tenant's requirements, the new offer should have a higher total price, lower unit prices, and more resources.

According to the amount of information provided by the client in the negotiation, the scenario can be subdivided into the following three cases:

\textbf{Case 1:} In each round, after receiving the new offer from the provider, the client only replies whether to accept it or not.

\textbf{Case 2:} In each round, after receiving the new offer from the provider, the client replies whether to accept it or not and the advice (which resource needs to be reduced).

\textbf{Case 3:} At the very beginning of the negotiation, the client tells the provider the priorities of the various resources (VCPU, RAM, and storage). 
In each round, after receiving the new offer from the provider, the client replies with whether to accept it or not and the same advice as in Case 2.

\begin{table}[!t]
\centering
\setlength{\tabcolsep}{3pt}
\customTableFont
\caption{\centering Simplified daily rental prices}
\begin{tabular}{c|c|c|c|c|c}
\toprule
\multicolumn{2}{c|}{\textbf{VCPU}}&\multicolumn{2}{|c|}{\textbf{RAM}}&\multicolumn{2}{|c}{\textbf{Storage}}\\ 
\midrule
Number  & €(per VCPU) &Number  & €(per GB)&  Number  & €(per GB)\\
\midrule
0-10 & level-1: 0.2 &  0-20 & level-1: 0.1 & 0-100&  level-1: 0.02 \\
11-30 & level-2: 0.1 & 21-60  &  level-2: 0.05 & 101-300 &  level-2: 0.01 \\
31-90 & level-3: 0.05  & 61-180 &  level-3: 0.025& 301-900  &  level-3: 0.005\\
\bottomrule
\end{tabular}
\label{tab:price}
\end{table}

%-------------------------------------------------------------------------------
\subsection{Core components}
%-------------------------------------------------------------------------------

\paragraph*{\textbf{Fuzzy negotiation system}}
We design and implement a fuzzy negotiation system (FNS) based on a one-to-one auto-negotiation model. The parties involved are two agents representing \emph{Provider} and \emph{Client}. Both agents are assumed to be rational beings \cite{osborne1994course}\cite{russell2010artificial}. The client's original requirement is used as the input of FNS. The final offer obtained by the auto-negotiation process is considered as the output of FNS. 
%\autoref{tab:dataset} shows a demo of a pair of input and output values.

A typical architecture of fuzzy logic consists of four parts: rule base, fuzzification, inference engine, and defuzzification. 
The fuzzy logic components are based on psychological and economic factors common in negotiation. 
Each part of the fuzzy system can be checked and understood by a human being, which is the source of interpretability \cite{babuvska2012fuzzy}.

Based on the above-mentioned, the working process of FNS can be described as: the client submits the original requirement as the input of FNS. 
The final offer obtained through the auto-negotiation process is used as the output of FNS. 
During each round of negotiation, the provider's agent updates the offer through specific algorithms, and the client's agent uses fuzzy logic to determine whether the updated offer is acceptable.

\paragraph*{\textbf{Surrogate modeling}}
Given a specific client, for any input, the FNS could generate a corresponding output. 
The way of using FNS is quite similar to that of machine learning models. On the other side, in previous studies \cite{abiodun2018state}\cite{wang2003artificial}\cite{hill1994artificial}, the neural network model has shown us that it works well in simulating human decision-making. 
We conjecture that the neural network should also have good performance in fitting the FNS that simulates the human way of thinking. 
The study by Carbo et al. \cite{carbo2003machine} partially verified our conjecture. Thus, we explore the feasibility of replacing the FNS with neural network models. 
We emphasize \emph{entire} to distinguish it from previous studies that used machine learning models to evaluate the offer in a single round.

Training a neural network model can be briefly described as follows: one input (client's original requirement) and its corresponding output (final offer) from the FNS are combined and considered a data record. 
The input is the \emph{feature}, and the output is the \emph{label} during the training process. 

A dataset could be obtained using different inputs and recording them and their corresponding outputs. Then we split the dataset into a test set and a training set to train the neural network models. 
The target of training is to make the output of the FNS and the model as close as possible.

Although auto-negotiation greatly speeds up the negotiation, each negotiation still requires dozens of rounds. 
In particular, in fuzzy logic, if the number of fuzzy rules is increased, the processing speed will be significantly reduced. Based on the above, we consider using both the fuzzy negotiation system and the neural network model. 
The fuzzy negotiation system provides interpretability and anthropomorphism. 
Then we use a black-box neural network model to replace the white-box fuzzy negotiation system, which greatly improves efficiency. 
Since the output of a machine learning model is similar to the fuzzy negotiation system, interpretability is preserved throughout the workflow.

\section{System architecture}\label{sec.systemdescription}
\subsection{Overall design}
\begin{figure}[!t]
\centerline{\includegraphics[width=\columnwidth]{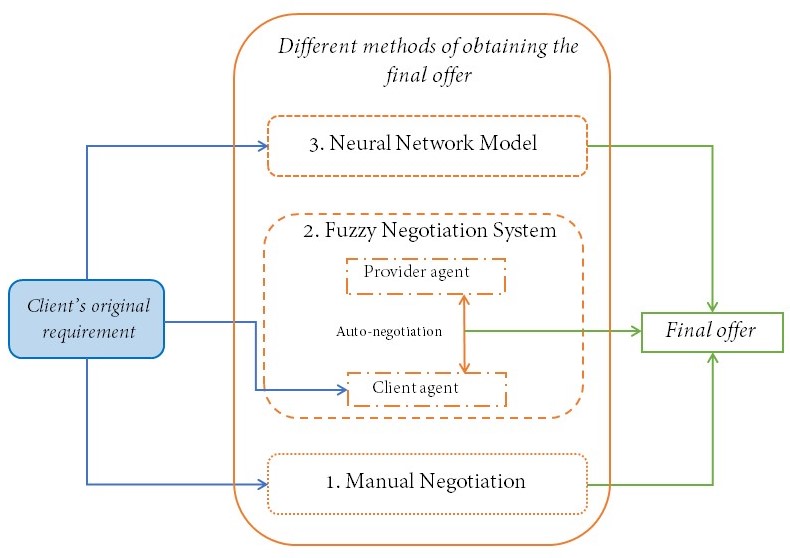}}
\caption{Overview of different methods}
\label{fig:systemfigure}
\end{figure}

\autoref{fig:systemfigure} is an overview of the methods involved in the article. 
It is unrealistic to rely entirely on manual negotiation. 
Therefore, we propose two other methods. The inputs of the three methods are consistent. 
Ideally, the output of the other two methods should be close to manual negotiation and they could speed up or partially replace manual negotiation.

%-------------------------------------------------------------------------------
\subsection{Fuzzy negotiation system}
%-------------------------------------------------------------------------------

%-------------------------------------------------------------------------------
\subsubsection{System workflow}
%-------------------------------------------------------------------------------

\begin{figure*}[!t]
\centering
\begin{subfigure}{1\columnwidth}
\centering
\includegraphics[trim=0pt 0pt 0pt 0pt, clip, width=0.8\columnwidth]{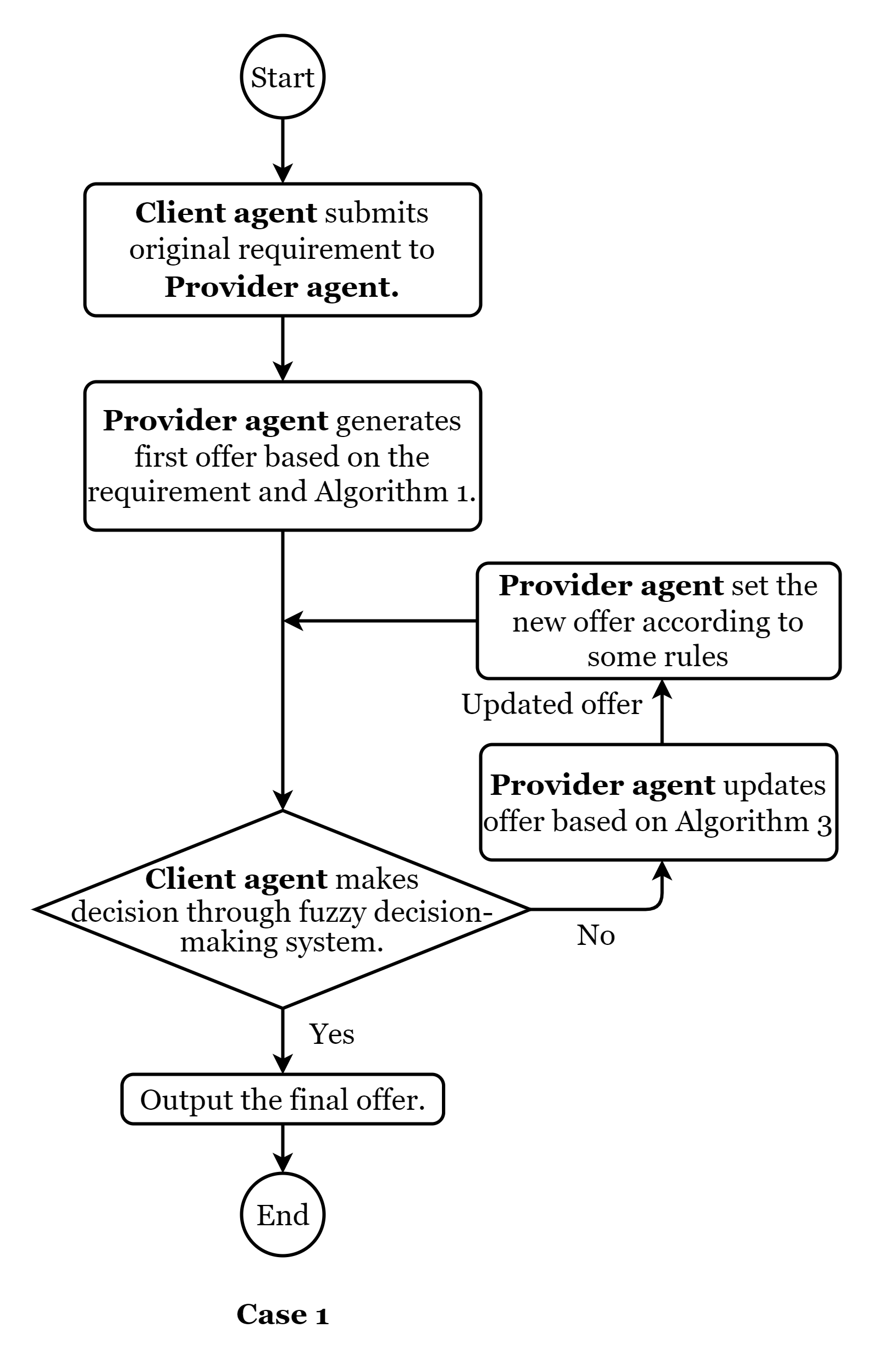}
\subcaption{The workflow in Case 1.}
\label{fig_case1}
\end{subfigure}
\begin{subfigure}{1\columnwidth}
\centering
\includegraphics[trim=0pt 0pt 0pt 0pt, clip, width=0.9\columnwidth]{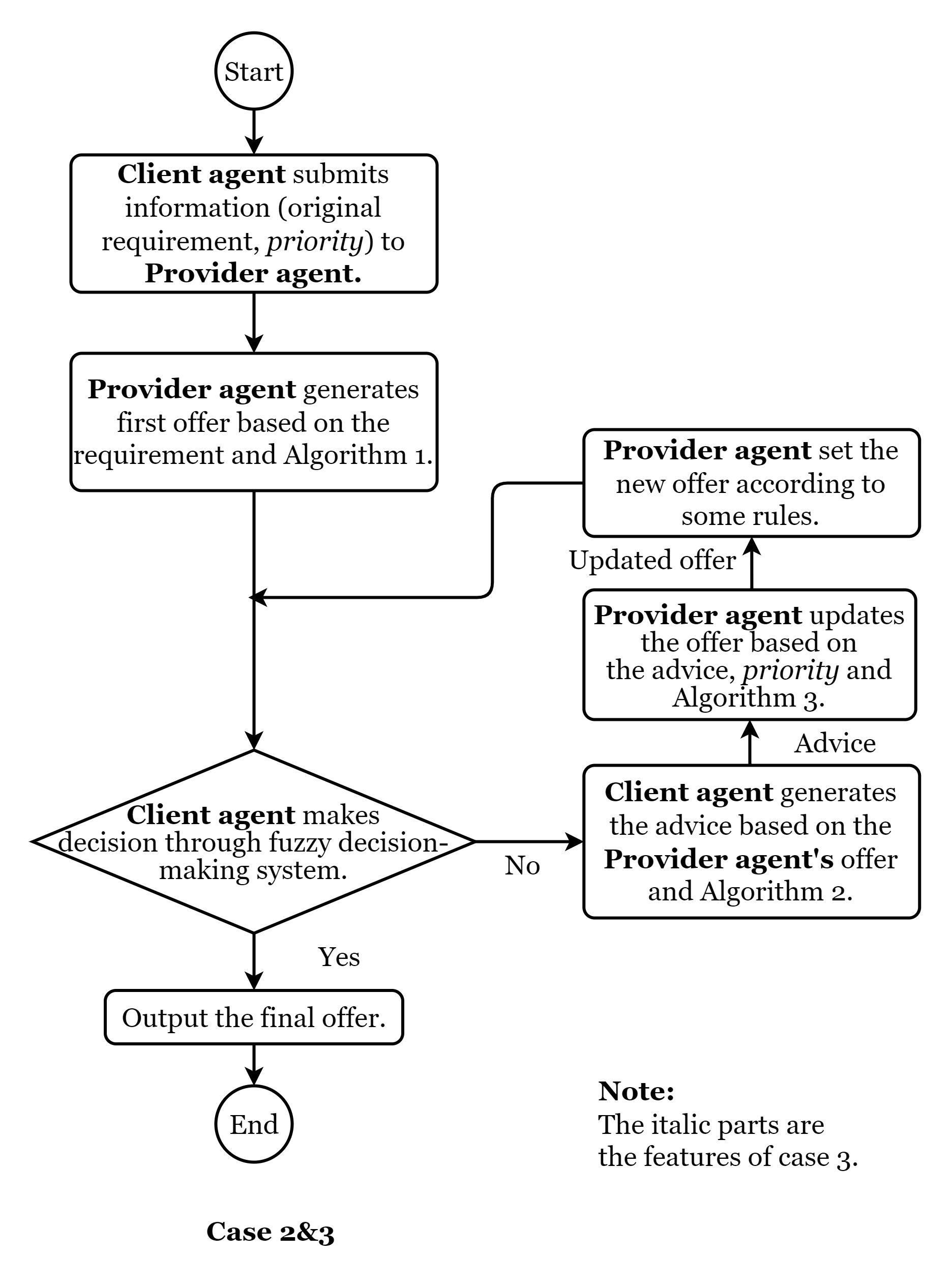}
\subcaption{The workflow in Case 2 and 3. The introduction of advice and priority is an important feature that distinguishes Case 1.}
\label{fig_case23}
\end{subfigure}
\caption{Flow charts in different cases. Algorithm 3 has multiple versions according to different cases.}
\label{fig:wkflow}
\end{figure*}

We use a direct-negotiation model in our fuzzy negotiation system. 
Different from indirect negotiation, two agents exchange information directly instead of submitting information to an intermediary. Therefore, the entire structure is simplified. 
The core parts are related algorithms and the fuzzy decision-making system. \autoref{fig:wkflow} shows the specific flow charts.

%-------------------------------------------------------------------------------
\subsubsection{Fuzzy decision-making system}
%-------------------------------------------------------------------------------

\paragraph*{Parameters} 
There are two types of parameters: crisp parameters and fuzzy parameters. 
Subscription plans from companies like Google, and Spotify, and some past studies \cite{nauges2000privately} \cite{scott2011matters} have shown that unit price and total order price are key factors in negotiation. 
In addition, Kristensen et al. \cite{kristensen1997adoption} show that in both the roles of buyers and sellers, subjects adopt an initial offer as a reference point. 
After comprehensively considering the above factors, we design four ratios ($ru_{c/0}^{VCPU}$, $ru_{c/0}^{RAM}$, $ru_{c/0}^{Storage}$, $rt_{0/c}$) to be the input of the system and use the client's tendency as the output. We set the client agent's original requirement as the reference point. \autoref{tab:crisp} shows the definition of crisp parameters. 
All crisp inputs are between 0 and 1, so the dimension effect is eliminated. 
Fuzzy values can be obtained by fuzzing crisp inputs using membership functions. 
The possible fuzzy values of each input and output of the fuzzy system are organized in \autoref{tab:fuzzy}. 
While describing the parameters, once again we would like to highlight that the choice of these particular parameters is based on the way current service providers offer services. The presented framework is not limited to only these parameters. 

\begin{table}[!t]
\begin{minipage}[!t]{\columnwidth}
\renewcommand{\thetable}{2(a)}
\caption{Definition of crisp parameters}
\label{tab:crisp}
\centering
\setlength{\tabcolsep}{3pt}
\customTableFont
\begin{tabular}{c|c}
\toprule
\multicolumn{1}{c|}{\textbf{Crisp parameters}}&\multicolumn{1}{c}{\textbf{Definition}}\\ 
\midrule
$pu_{0}^{VCPU}, pu_{0}^{RAM}, pu_{0}^{Storage}$ & Resources' unit prices in original requirements \\
\midrule
$pu_{c}^{VCPU}, pu_{c}^{RAM}, pu_{c}^{Storage}$ & Resources' unit prices in current offers \\
\midrule
$pt_{0}$ &  Total price of the original requirement \\ 
\midrule
$pt_{c}$ &  Total price of the current offer \\ 
\midrule
$ru_{c/0}^{VCPU}$  & $ru_{c/0}^{VCPU} = pu_{c}^{VCPU}/pu_{0}^{VCPU}$\\
\midrule
$ru_{c/0}^{RAM}$ & $ru_{c/0}^{RAM} = pu_{c}^{RAM}/pu_{0}^{RAM}$\\
\midrule
$ru_{c/0}^{Storage}$ & $ru_{c/0}^{Storage} = pu_{c}^{Storage}/pu_{0}^{Storage}$\\
\midrule
$rt_{0/c}$ & $rt_{0/c} = pt_{0}/pt_{c}$ \\
\bottomrule
\end{tabular}
\end{minipage}
\\[12pt] 
\begin{minipage}[!t]{\columnwidth}
\renewcommand{\thetable}{2(b)}
\caption{Possible values of fuzzy parameters}
\label{tab:fuzzy}
\centering
\setlength{\tabcolsep}{3pt}
\customTableFont
\begin{tabular}{c|c|c|c}
\toprule
\multicolumn{1}{c|}{\textbf{Input}}&\multicolumn{3}{c}{Corresponding fuzzy values}\\ 
\midrule
VUPR (VCPU unit price ratio, current/original)  & cheap & medium  & expensive\\
\midrule
RUPR (RAM unit price ratio, current/original) & cheap & medium  & expensive\\
\midrule
SUPR (Storage unit price ratio, current/original) & cheap & medium  & expensive\\
\midrule
TPR (Total price ratio, original/current) & high & medium  & low\\
\midrule
\multicolumn{1}{c|}{\textbf{Output}}&\multicolumn{3}{c}{Corresponding fuzzy values}\\ 
\midrule
Tendency  & low & medium  & high \\
\bottomrule
\end{tabular}
\end{minipage}
\setcounter{table}{2}
\end{table}

\paragraph*{Membership functions} 
We choose two types of membership functions: the triangular membership function and the Gaussian membership function. In the same system, we use only one type of membership function instead of mixing two types. For Gaussian membership functions, we apply the 3-sigma rule \cite{pukelsheim1994three}. The starting point, the highest point, and the ending point of the triangular function are obtained at $a$, $b$, and $c$, respectively. The mean of the corresponding Gaussian function is $m$ and the standard deviation is $\sigma$, then $m=b, a=b-3\sigma, c=b+3\sigma$. The three ratios share the same membership functions while the tendency uses a unique one. 

\paragraph*{Fuzzy rule set} 
We design two fuzzy rule sets based on prior knowledge: a 7-rule set and a 30-rule set. 
The 7-rule set refers to \autoref{tab:7rules}.
Complex rules lead to better anthropomorphism, but performance decreases. 
We conduct relevant experiments to compare and analyze the two rule sets. 
Fuzzy outputs could be obtained according to fuzzy rule sets. 
\paragraph*{Defuzzification} It converts fuzzy sets obtained by the inference engine into crisp values. Usually, a specific defuzzification method is selected and combined with a specific expert system to improve accuracy. We apply the Center of Gravity (CoG) method \cite{van2004comparison} during defuzzification to transfer fuzzy inference results into crisp outputs (tendency score).
\paragraph*{Threshold of the final scores} The threshold is used to evaluate whether the offer is accepted. According to \autoref{tab:price}, for crisp input parameters, the unit price ratio is in the range [0.389, 1], and the ratio of the total price is in the range [0.0238, 1]. So in the worst case (VUPR:1, RUPR:1, SUPR:1, TPR:0.0238), the final score is about 14.79. While in the best case (VUPR:0.389, RUPR:0.389, SUPR:0.389, TPR:1), the final score is about 61.63. In fact, the best and worst cases cannot happen, so the final score ranges between 14.79 and 61.63. For the original requirement, the score is 50.00. Given the above, the threshold should be set to a number close to 50.00 based on some prior knowledge.

\begin{table}[!t]
\centering
\setlength{\tabcolsep}{3pt}
\customTableFont
\caption{\centering The seven-rule set}
\begin{tabular}{c|c|c|c|c|c}
\toprule
\multicolumn{1}{c|}{}&\multicolumn{4}{c|}{Fuzzy input}& Fuzzy output\\ 
\midrule
Rule index & VUPR & RUPR & SUPR  & TPR & Tendency\\
\midrule
1 & $\backslash$ & $\backslash$ & $\backslash$  & high & low\\
\midrule
2 & $\backslash$ & $\backslash$ & $\backslash$  & medium & medium\\
\midrule
3 & $\backslash$ & $\backslash$ & $\backslash$  & low & high\\
\midrule
4 & cheap & cheap & cheap  & $\backslash$ & high\\
\midrule
5 & expensive & $\backslash$ & $\backslash$  & $\backslash$ & medium\\
\midrule
6 & $\backslash$ & expensive & $\backslash$  & $\backslash$ & medium\\
\midrule
7 & $\backslash$ & $\backslash$ & expensive  & $\backslash$ & medium\\
\bottomrule
\end{tabular}
\label{tab:7rules}
\end{table}

%-------------------------------------------------------------------------------
\subsubsection{Algorithms}
%-------------------------------------------------------------------------------

\paragraph*{Algorithm 1}
\label{p:algorithm1} 
Both the client agent's original requirements and the provider agent's offers consist of the quantities of VCPU, RAM, and storage. 
The motivation for clients to rent more resources is lower unit prices. 
Therefore, when the provider agent generates the first offer, the unit price of all parameters in it should be lower than the client agent's original requirement. 
To achieve this goal, the quantity of each parameter needs to be increased by $p$ so that a cheaper unit price can be adopted. 
Thresholds for different levels of prices determine $p$.
For example, if the number of VCPU is increased from 5 to 15, the unit price used is level 2 instead of level 1. 
The details are available in \autoref{alg:1} in the appendix.

\paragraph*{Algorithm 2}
\label{p:algorithm2} 
During negotiation, if the client agent rejects the offer, it could advise the provider agent on how to update the offer. In real life, clients often require the quantities of different resources to be in a specific ratio, that is:

\begin{footnotesize}
\begin{equation}
    \label{eq:1}
    \frac {n_c^{VCPU}} {n_0^{VCPU}}=\frac {n_c^{RAM}} {n_0^{RAM}}=\frac {n_c^{Storage}} {n_0^{Storage}}
\end{equation}
\end{footnotesize}
\autoref{eq:1} could be rewritten as follows:
\begin{footnotesize}
\begin{equation}
    \label{eq:1b}
    \frac {max(\frac {n_c^{VCPU}} {n_0^{VCPU}},\frac {n_c^{RAM}} {n_0^{RAM}},\frac {n_c^{Storage}} {n_0^{Storage}})}{min(\frac {n_c^{VCPU}} {n_0^{VCPU}},\frac {n_c^{RAM}} {n_0^{RAM}},\frac {n_c^{Storage}} {n_0^{Storage}})} = 1
\end{equation}
\end{footnotesize}

where:

\begin{footnotesize}
\begin{conditions}
 n_0^{VCPU},n_0^{RAM},n_0^{Storage} &  the quantities of VCPU, RAM, and storage in the original requirement \\
 n_c^{VCPU},n_c^{RAM},n_c^{Storage} &  the quantities of VCPU, RAM, and storage in the current offer \\ 
\end{conditions}
\end{footnotesize}

For the negotiation to continue, we slightly relax the conditions of \autoref{eq:1b}, namely:

\begin{footnotesize}
\begin{equation}
    \label{eq:2}
    \frac {max(\frac {n_c^{VCPU}} {n_0^{VCPU}},\frac {n_c^{RAM}} {n_0^{RAM}},\frac {n_c^{Storage}} {n_0^{Storage}})}{min(\frac {n_c^{VCPU}} {n_0^{VCPU}},\frac {n_c^{RAM}} {n_0^{RAM}},\frac {n_c^{Storage}} {n_0^{Storage}})} < d_{max}
\end{equation}
\end{footnotesize}

where:

\begin{footnotesize}
\begin{conditions}
d_{max} & the maximum deviation value, usually slightly larger than 1, specified by the client  
\end{conditions}
\end{footnotesize}

If \autoref{eq:2} is satisfied, the client agent gives the advice \emph{random}. Otherwise, the client agent uses the parameter with the largest quantity as the advice. The details are shown in \autoref{alg:2} in the appendix.

\paragraph*{Algorithm 3}
\label{p:algorithm3} 
To speed up the negotiation, a search problem, a method that draws on the idea of stochastic gradient descent is applied. In each round, we randomly select one of the three parameters in the offer to reduce with some probability. Given the amount of changing $c^{Para}$ in each parameter per round, the selected parameter will be reduced by $c^{Para}$ to update the offer. In different cases, we have slightly different algorithms for calculating the probabilities and selecting the parameters. 

\paragraph*{Case a}
\label{par:caseA}
At a given round of negotiations, the original requirement, current offer, and final offer satisfy:

\begin{footnotesize}
\begin{gather}
    \label{eq:3}
     n_c^{Para} \approx q \cdot n_0^{Para}, \quad 1 \leq q,\quad q \in \mathbb{R}\\
     n_f^{Para} \approx k \cdot n_0^{Para}, \quad 1 \leq k \leq q,\quad k \in \mathbb{R}
\end{gather}
\end{footnotesize}

where:

\begin{footnotesize}
\begin{conditions}
 n_c^{Para} &  the quantity of a specific parameter in the current offer \\ 
 n_0^{Para} &  the quantity of the same parameter in the original requirement \\
 n_f^{Para} &  the quantity of the same parameter in the final offer \\ 
\end{conditions}
\end{footnotesize}

From the current offer to the final offer, the number of rounds required can be expressed as:

\begin{footnotesize}
\begin{gather}
    \label{eq:4}
     rd^{Para}=(q-k) \cdot n_0^{Para}/c^{Para}
\end{gather}
\end{footnotesize}

where:

\begin{footnotesize}
\begin{conditions}
c^{Para} & the amount of change of the same parameter per round
\end{conditions}
\end{footnotesize}

In this case, we use the ratio of the number of rounds needed for each parameter to represent the ratio of the probabilities corresponding to each parameter. 
With the assumption that the ratios remain consistent with the initial ratios in the original requirement, all the parameters should share the same $q$ and $k$ at the same round. Then the probabilities satisfy:

\begin{footnotesize}
\begin{gather}
\begin{split}
    \label{eq:5}
       &\ pb^{VCPU}:pb^{RAM}:pb^{Storage}\\
     =&\ rd^{VCPU}:rd^{RAM}:rd^{Storage}\\
     =&\ \frac {n_0^{VCPU}} {c^{VCPU}}:\frac {n_0^{RAM}} {c^{RAM}}:\frac {n_0^{Storage}} {c^{Storage}}
\end{split}
\end{gather}
\end{footnotesize}

where:

\begin{footnotesize}
\begin{conditions}
pb^{VCPU}, pb^{RAM}, pb^{Storage} & the probability of choosing the parameter in a round
\end{conditions}
\end{footnotesize}

The details can be found in \autoref{alg:3a} in the appendix.

\paragraph*{Case b}
\label{par:caseB}
The ratio of the number of parameters in the offer may deviate a lot from the client's requirement during the negotiation. 
At this time, with \autoref{alg:3a}, it may not be possible to obtain an offer that meets a specific proportion. 
Therefore, we introduce \emph{Advice} on the client side to help the provider to revise the offer by \autoref{alg:2}.
If the \emph{Advice} given by the client is 'randomly', parameters requiring a reduction in amount are selected according to \autoref{alg:3a}. 
If the \emph{Advice} is the name of a certain parameter, the number of that parameter will be reduced.
Details can be found in \autoref{alg:3b} in the appendix.

\paragraph*{Case c}
\label{par:caseC}
In real life, the client thinks that different parameters have different priorities. 
This means that when a customer considers both offers acceptable, the customer will prefer the offer with higher high-priority parameters. 
We introduce the parameter \emph{priority} to modify the probability so that the offer could fit the client's requirement more. 
Each parameter has its respective \emph{priority}. 
The value of \emph{priority} is greater than or equal to 1. A parameter with a higher priority has a larger value of \emph{priority}. 
The algorithm using adjusted probabilities is described in \autoref{alg:3c} in the appendix.

%-------------------------------------------------------------------------------
\subsection{Machine learning process}
%-------------------------------------------------------------------------------

%-------------------------------------------------------------------------------
\subsubsection{Workflow}
%-------------------------------------------------------------------------------

\begin{figure}[!t]
\centerline{\includegraphics[width=1\columnwidth]{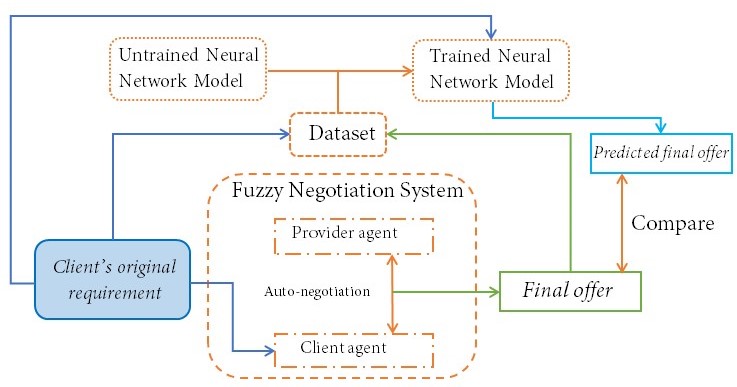}}
\caption{Workflow of training machine learning models}
\label{fig:mlwf}
\end{figure}

We aim to replace the entire fuzzy negotiation system with a neural network surrogate model. 
We use fuzzy negotiation systems to generate the training and test datasets: each original requirement and its corresponding output of the fuzzy negotiation system is treated as a record and stored in the database.
These data are normalized. 
We treat this process as a regression problem. The original requirement is considered as the input of models while the output of the fuzzy system (a vector consisting of the amounts of three sources) is considered as the label. The mean squared error (MSE) is used to measure the distance (loss) between the label vectors and the model's output vectors.
After training, we input the same original requirements into the trained model and the fuzzy negotiation system, respectively, and compare the outputs of the two systems. 
Refer to \autoref{fig:mlwf} for the entire workflow.

%-------------------------------------------------------------------------------
\subsubsection{Model architecture}
%-------------------------------------------------------------------------------

For regression problems, activation functions (e.g., Softmax, Sigmoid, etc.) are not used in the last layer of the neural network model. We try two types of models with different structures and layers. 

Type 1 is the multilayer perceptron (MLP). 
It consists of four linear layers. 
It expands the input three features to 128 dimensions and then scales to 64 dimensions and 27 dimensions, and the final output contains three dimensions. 
Furthermore, we change the number of layers of the model to analyze the effect of the depth of the neural network model.

Type 2 uses a 7-layer architecture. Among them, the fourth layer is the Relu layer, and the others are linear layers. 
The role of the Relu layer is to introduce nonlinear components. It is generally believed that fuzzy logic is suitable for nonlinear systems. 
So we conjecture Type 2 will have better performance than Type 1.

%-------------------------------------------------------------------------------
\section{Results and discussion} 
\label{sec.exp}
%-------------------------------------------------------------------------------

%-------------------------------------------------------------------------------
\subsection{Settings} 
\label{sec:setexp}
%-------------------------------------------------------------------------------

We use three sizes of input datasets to analyze the results or performance. Details of these datasets can be found in \autoref{tab:settingdata}. \autoref{tab:systemsettings} shows the information on the different systems we use. The cases are defined in \autoref{Scenario}. The experiments are based on \textit{Python} and its library \textit{Scikit-fuzzy}. The associated codes and datasets can be found in the open source  \href{https://github.com/Junjie-Chu/Efficient_Resource_Scheduling_for_Distributed_Infrastructures_using_Negotiation_Capabilities}{repository}.

\begin{table}[!thbp]
\centering
\setlength{\tabcolsep}{3pt}
\customTableFont
\caption{\centering Different input datasets used in the experiment}
\begin{tabular}{c|c|c}
\toprule
Index & No. records & Note \\
\midrule
Dataset 1 & 1 & The input is (10,20,200) \\
\midrule
Dataset 2 & 200 & Test dataset, total fee from low to high.\\
\midrule
Dataset 3 & 10000 & Training dataset, No overlap with Dataset 2.\\
\bottomrule
\end{tabular}
\label{tab:settingdata}
\end{table}

\begin{table}[!thb]
\centering
\setlength{\tabcolsep}{3pt}
\customTableFont
\caption{\centering Configuration of different systems}
\begin{tabular}{c|c|c|c|c}
\toprule
Version & Case & Membership function & Fuzzy rule set & Threshold (final score) \\
\midrule
System 1 & Case 2 & Triangular & 7-rule set & 50.0\\
\midrule
System 2 & Case 2 & Gaussian & 7-rule set & 50.0\\
\midrule
System 3 & Case 2 & Gaussian & 30-rule set & 50.0\\
\midrule
System 4 & Case 1 & Gaussian & 30-rule set & 50.0\\
\midrule
System 5 & Case 3 & Gaussian & 30-rule set & 50.0\\
\bottomrule
\end{tabular}
\label{tab:systemsettings}
\end{table}

%-------------------------------------------------------------------------------
\subsection{Experiments of fuzzy negotiation system} 
\label{sec:fuzzyexperiment}
%-------------------------------------------------------------------------------

%-------------------------------------------------------------------------------
\subsubsection{Different membership functions} 
\label{sec:dfmf}
%-------------------------------------------------------------------------------

In this experiment, all parameters are the same except for the membership functions used in the client agent. We compare System 1 and System 2 in \autoref{tab:systemsettings} on Dataset 2.

The ratios of the total fee (final offer: original requirement) are shown in \autoref{fig.diffmf0}. If the ratio is 1.0, the negotiation fails. 
The negotiation success rates of versions with different membership functions are almost the same (61.5\%). 
However, compared with the triangular membership function, the total fees of the final offers output by the system with the Gaussian membership function are higher, and the fees fluctuate more. 
We observe similar phenomena with the amounts of other parameters. 
Therefore, we believe the client agent using the Gaussian membership function is more flexible.

\begin{table}[!htb]
\centering
\setlength{\tabcolsep}{3pt}
\customTableFont
\caption{\centering Performance of different membership functions}
\begin{tabular}{c|c|c|c|c}
\toprule
Version & Dataset & Records number & Rounds number & Time \\
\midrule
System 1 (Triangular)& Dataset 1& 1 & 42 & 1.316s\\
\midrule
System 1 (Triangular)& Dataset 2& 200 & $\backslash$ & 1m12s\\
\midrule
System 1 (Triangular)& Dataset 3& 10000 & $\backslash$ & 1h6m24s\\
\midrule
System 2 (Gaussion)& Dataset 1 & 1 & 27 & 0.051s\\
\midrule
System 2 (Gaussion)& Dataset 2 & 200 & $\backslash$ & 3.012s\\
\midrule
System 2 (Gaussion)& Dataset 3 & 10000 & $\backslash$ & 2m56s\\
\bottomrule
\end{tabular}
\label{tab:diffmftime}
\end{table}

In addition, we test the performance of the two versions on three datasets. 
\autoref{tab:diffmftime} shows the results. 
The version of the Gaussian membership function has better performance on all datasets. 
The number of negotiation rounds between the two versions is not much different (42 vs. 27) on Dataset 1. 
But the time-consuming difference is dozens of times (1.316s vs. 0.051s). 
We draw similar conclusions from other datasets.

%-------------------------------------------------------------------------------
\subsubsection{Different fuzzy rule sets} 
\label{sec:dfrule}
%-------------------------------------------------------------------------------

In this experiment, we test System 2 and System 3 in \autoref{tab:systemsettings} on Dataset 2. The difference between them lies in different fuzzy rule sets. 

The output trends of the two versions are similar as shown in \autoref{fig.diffrule0}. 
Compared with the 7-rule version, the total fee ratio of the 30-rule version fluctuates more, which means that the 30-rule version simulates the thinking of a real human being to a higher degree. 
The 30-rule version is slightly lower than the 7-rule version in success rate (60\% vs. 61.5\%). 

\begin{table}[!t]
\centering
\setlength{\tabcolsep}{3pt}
\customTableFont
\caption{\centering Performance test of different fuzzy rule sets}
\begin{tabular}{c|c|c|c|c}
\toprule
Version & Dataset & Number of records & Number of rounds & Time \\
\midrule
System 2 (7-rule)& Dataset 1& 1 & 27 & 0.051s\\
\midrule
System 2 (7-rule)& Dataset 2& 200 & $\backslash$ & 3.012s\\
\midrule
System 2 (7-rule)& Dataset 3& 10000 & $\backslash$ & 2m56s\\
\midrule
System 3 (30-rule)& Dataset 1 & 1 & 55 & 0.342s\\
\midrule
System 3 (30-rule)& Dataset 2 & 200 & $\backslash$ & 48.011s\\
\midrule
System 3 (30-rule)& Dataset 3 & 10000 & $\backslash$ & 39m04s\\
\bottomrule
\end{tabular}
\label{tab:diffruletime}
\end{table}

Different fuzzy rule sets also result in different performances. As shown in \autoref{tab:diffruletime}, the 30-rule system spends much longer time on all datasets than the 7-rule system. From a time-consuming perspective, the 7-rule version has better performance. The curve of the 30-rule system is more complex than that of the 7-rule system, reflecting that the 30-rule system can be better anthropomorphic. From the perspective of simulating human thinking, it is worth mentioning that the 30-rule system using the Gaussian membership function performs better than the 7-rule one using the Gaussian membership function.

%-------------------------------------------------------------------------------
\subsubsection{Different cases} 
\label{sec:dfcase}
%-------------------------------------------------------------------------------

We conduct experiments to compare the systems under different cases. The systems we use are System 3, System 4, and System 5 in \autoref{tab:systemsettings}.

\begin{figure*}[!t]
\centering
\begin{subfigure}{1\columnwidth}
\centering
\includegraphics[trim=0pt 0pt 0pt 0pt, clip, width=0.95\columnwidth]{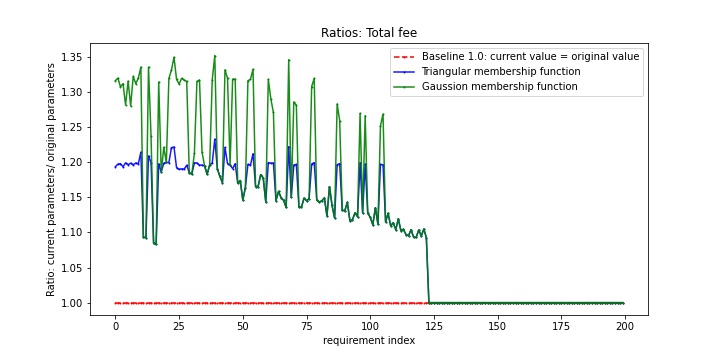}
\subcaption{Ratios of total fees in Dataset 2 (different membership functions)}
\label{fig.diffmf0}
\end{subfigure}
\begin{subfigure}{1\columnwidth}
\centering
\includegraphics[trim=0pt 0pt 0pt 0pt, clip, width=0.95\columnwidth]{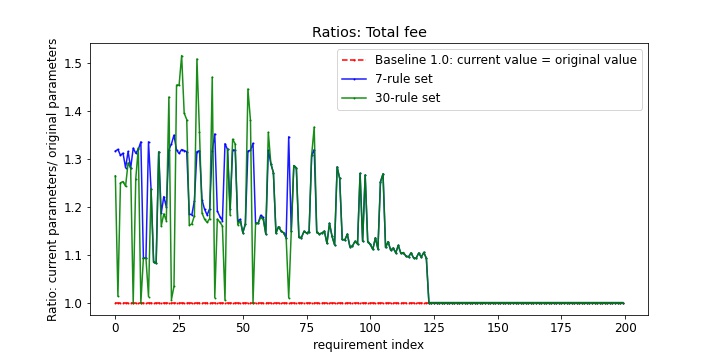}
\subcaption{Ratios of total fees in Dataset 2 (different rule sets)}
\label{fig.diffrule0}
\end{subfigure}
\begin{subfigure}{1\columnwidth}
\centering
\includegraphics[trim=0pt 0pt 0pt 0pt, clip, width=0.95\columnwidth]{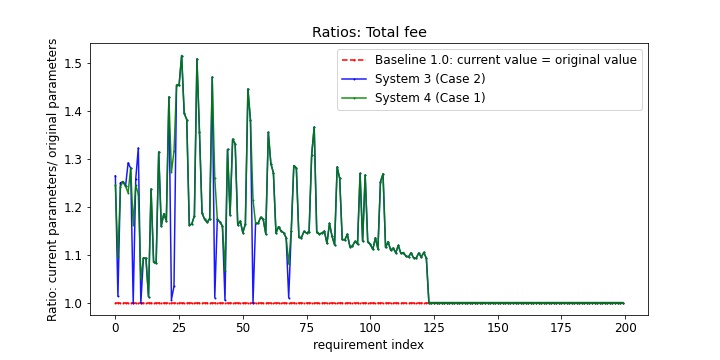}
\subcaption{Ratios of total fees in Dataset 2 (different cases)}
\label{fig.diffcase0}
\end{subfigure}
\begin{subfigure}{1\columnwidth}
\centering
\includegraphics[trim=0pt 0pt 0pt 0pt, clip, width=0.95\columnwidth]{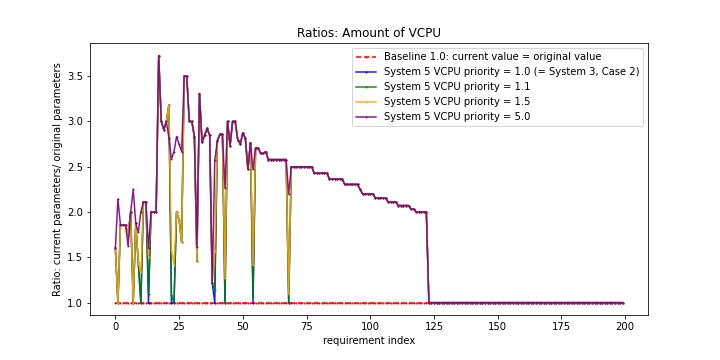}
\subcaption{Ratios of VCPU's amounts on Dataset 2 (different priorities)}
\label{fig.diffpriocpu}
\end{subfigure}
\caption{Plots of ratios in different experiments. 
Horizontal axes represent the index of the input requirement. 
The vertical axes represent ratios. 
The baseline is 1.0, which means values in final offers are the same as those in the original requirements.}
\label{Exp_Nego_Figures}
\end{figure*}

\paragraph{Case 1 vs. Case 2}
We conduct related experiments on Dataset 2. According to \autoref{fig.diffcase0}, the results of these two cases are generally similar. When the number of leased resources is large (level-3 prices apply), there is no room for negotiation, so the ratio is always 1.0 (negotiation fails). When the number of rented resources is relatively tiny (level-1 or level-2 price applies), the negotiation success rate of Case 2 is slightly lower than that of Case 1. Usually, the total fee ratio of Case 2 is also a bit lower than that of Case 1. This aligns with our expectation: Case 1 only considers the client's financial factors. Case 2 adds requirements for the ratio of each parameter on this basis, so the negotiation of Case 2 is more complicated.

\paragraph{Case 2 vs. Case 3}
In Case 3, we introduce $priority$ to modify the final offer. 
Case 2 is a special case of Case 3 when all parameters have the same priority. 
As shown in \autoref{tab:diffprio}, the system of Case 3 tends to select offers with more high-priority parameters among all the offers that satisfy other conditions. 
We fix the $priority$ values of RAM and storage (set to 1), change the $priority$ value of VCPU, and conduct experiments on Dataset 2. 
As shown in \autoref{fig.diffpriocpu}, the conclusion is similar to that of the above single-input experiment. 
The version with the highest VCPU $priority$ value retains the most VCPUs. 

\paragraph{Performance}
The features we add in case 2 and case 3 aim to make the offer more in line with the client's requirements rather than reducing the number of negotiation rounds. In other words, the parameters given by $advice$ or $priority$ do not necessarily result in the fastest gradient descent direction. Therefore, the performance of Case 2 or 3 is not necessarily better than that of Case 1. As shown in \autoref{tab:diffcasetime}, the performance of the three cases is close. The time to process 10,000 entries is typically 40 minutes or more. So it is necessary to look for a more efficient alternative.

\begin{figure}[!b]
\centering
\includegraphics[width=0.9\columnwidth]{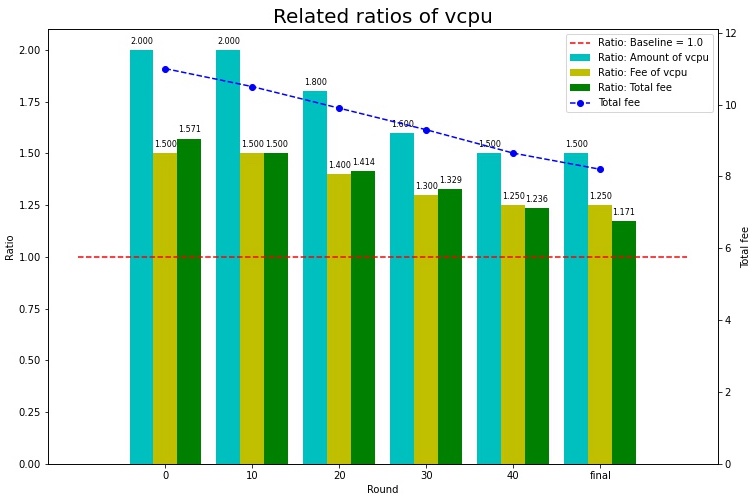}
\caption{Comparison of the ratios (VCPU)}
\label{fig.ratiocpu}   
\end{figure}

%-------------------------------------------------------------------------------
\subsubsection{Evaluation of the negotiation results}
\label{sec:evaluation}
%-------------------------------------------------------------------------------

We test three different versions of the system on Dataset 2. In System 5, the $priority$ value of VCPU is set to 1.50, and those of other parameters are set to 1.0. We evaluate the systems based on the total fee ratios and negotiation success rate. A higher ratio means the provider rents more resources and earns more profit.

As shown in \autoref{tab:avgfeeratio}, over the entire Dataset 2, the total fee ratios of all versions increase by more than 10\%. We remove the elements in Dataset 2 that have no negotiation space (that is, the unit price of the resource is already level 3) to get a new dataset. The total fee ratio improvement on the new dataset is over 15\% for each version. In terms of negotiation success rate, the success rate on the new dataset is very high. Among them, the least restrictive version (System 4) has the highest negotiation success rate, reaching 100\%. Providers can earn at least an average of 15\% extra in the vast majority of cases where there is room for negotiation.

Comparing the different ratios in an offer makes the attractiveness of the offer more intuitive. 
We use System 5 and Dataset 1 for demonstration. 
\autoref{fig.ratiocpu} shows the results of VCPU and the total fee for some negotiation rounds. 
In the final offer (the 46th round), the client only needs to spend 17.1\% more to get 50\% more VCPU. 
The situation is similar for the other parameters: the client can also get 30\% more RAM and 20\% more storage.
From these ratios, it is clear that clients can get a lot of additional resources for only a small extra cost.

\begin{table}[!t]
\centering
\setlength{\tabcolsep}{3pt}
\customTableFont
\caption{Experiments of different priorities in Case 3. The priority of RAM and Storage is set to 1.0, and only the priority values of VCPU are changed. Inputs and outputs are in the format: (number of VCPUs, RAM, and Storage).}
\begin{tabular}{c|c|c|c|c}
\toprule
{Input}&{Priority of VCPU}&{Output} & {Score of output}& {Number of rounds}\\ 
\midrule
10, 20, 200 & 1.00 & 10, 23, 230 & 50.075 & 55\\
\midrule
10, 20, 200 & 1.10 & 12, 23, 220 & 50.123 & 54\\
\midrule
10, 20, 200 & 1.50 & 15, 26, 240 & 50.610 & 46\\
\midrule
10, 20, 200 & 5.00 & 17, 26, 260 & 50.442 & 42\\
\bottomrule
\end{tabular}
\label{tab:diffprio}
\end{table}

\begin{table}[t]
\centering
\setlength{\tabcolsep}{3pt}
\customTableFont
\caption{\centering Performance test of different cases}
\begin{tabular}{c|c|c|c|c}
\toprule
Version & Dataset & Number of records & Number of rounds & Time \\
\midrule
System 3 (case 2)& Dataset 1 & 1 & 55 & 0.342s\\
\midrule
System 3 (case 2)& Dataset 2 & 200 & $\backslash$ & 48.011s\\
\midrule
System 3 (case 2)& Dataset 3 & 10000 & $\backslash$ & 39m04s\\
\midrule
System 4 (case 1)& Dataset 1 & 1 & 27 & 0.254s\\
\midrule
System 4 (case 1)& Dataset 2 & 200 & $\backslash$ & 40.076s\\
\midrule
System 4 (case 1)& Dataset 3 & 10000 & $\backslash$ & 41m15s\\
\midrule
System 5 (case 3)& Dataset 1 & 1 & 45 & 0.311s\\
\midrule
System 5 (case 3)& Dataset 2 & 200 & $\backslash$ & 73.032s\\
\midrule
System 5 (case 3)& Dataset 3 & 10000 & $\backslash$ & 54m06s\\
\bottomrule
\end{tabular}
\label{tab:diffcasetime}
\vspace{-1.0em}
\end{table}

\begin{table}[t]
\centering
\setlength{\tabcolsep}{3pt}
\customTableFont
\caption{\centering Average total fee ratio}
\begin{tabular}{c|c|c|c|c}
\toprule
Version & Dataset & Num of records & Avg. total fee ratio & Success rate \\
\midrule
System 3 (Case 2)& Dataset 2 & 200 & 1.114 & 60.0\% \\
\midrule
System 4 (Case 1)& Dataset 2 & 200 & 1.120 & 61.5\% \\
\midrule
System 5 (Case 3)& Dataset 2 & 200 & 1.108 & 61.0\% \\
\midrule
System 3 (Case 2)& Dataset 2  & 124 & 1.185 & 97.6\%\\
\midrule
System 4 (Case 1)& Dataset 2  & 124 & 1.195 & 100.0\%\\
\midrule
System 5 (Case 3)& Dataset 2 & 124 & 1.176 & 99.2\%\\
\bottomrule
\end{tabular}
\label{tab:avgfeeratio}
\end{table}

%-------------------------------------------------------------------------------
\subsection{Experiments of machine learning} 
\label{sec:mlexperiment}
%-------------------------------------------------------------------------------

%-------------------------------------------------------------------------------
\subsubsection{Simulate the same system with different models}
\label{sec:diffmodel}
%-------------------------------------------------------------------------------

\begin{table}[!t]
\centering
\setlength{\tabcolsep}{3pt}
\customTableFont
\caption{\centering MSE loss of different models' predictions}
\begin{tabular}{c|c|c|c|c}
\toprule
Version &  VCPU & RAM & Storage & Entire offer\\
\midrule
Model 1 & 0.0159 & 0.0055 & 0.0060 & 0.0092 \\
\midrule
Model 2 & 0.0155 & 0.0057 & 0.0057 & 0.0090 \\
\midrule
Model 3 & 0.0157 & 0.0055 & 0.0057 & 0.0090 \\
\midrule
Model 4 & 0.0018 & 0.0009 & 0.0005 & 0.0011  \\
\bottomrule
\end{tabular}
\label{tab:modelresults}
\end{table}

\begin{table}[t]
\centering
\setlength{\tabcolsep}{3pt}
\customTableFont
\caption{\centering MSE loss of predictions on different systems}
\begin{tabular}{c|c|c|c|c}
\toprule
Version & VCPU & RAM & Storage & whole offer\\
\midrule
System 1 (Case 2) & 0.0009 & 0.0004 & 0.0005 & 0.0006 \\
\midrule
System 2 (Case 2)& 0.0007 & 0.0011 & 0.0003 & 0.0007 \\
\midrule
System 3 (Case 2)& 0.0011 & 0.0007 & 0.0006 & 0.0008 \\
\midrule
System 4 (Case 1)& 0.0008 & 0.0010 & 0.0006 & 0.0008  \\
\midrule
System 5 (Case 3)& 0.0011 & 0.0006 & 0.0006 & 0.0008  \\
\bottomrule
\end{tabular}
\label{tab:systemresults}
\end{table}

We use System 3 as the negotiation system to be replaced. 
Dataset 2 and its output in System 3 are combined as the test dataset. 
Dataset 3 and its output in System 3 are combined as a training dataset. 
The number of epochs is 200.
We test four models in total.
Models 1, 2, and 3 contain 4, 6, and 7 linear layers, respectively. 
The first three models consist of linear layers only. 
Model 4 consists of 6 linear layers and 1 Relu layer.
We separately evaluate the predicted quantity for a single parameter and the entire predicted offer (i.e., a vector of quantities of three items) with mean squared error (MSE). 
\autoref{tab:modelresults} presents these results. 
%\autoref{fig.comparea}, \autoref{fig.compareb}, \autoref{fig.comparec} in Appendix show the comparison of labels and predictions.

For models with only linear layers, the number of layers slightly improves prediction accuracy. 
The gap between the predictions and labels of these models indicates that these models are insufficient to replace the original negotiation system. 
However, the Relu layer significantly improves the performance of the model. 
The predictions of the model with the Relu layer are very close to the labels.

\begin{table}[!t]
\centering
\setlength{\tabcolsep}{3pt}
\customTableFont
\caption{\centering Number of VCPUs in intermediate rounds}
\begin{tabular}{c|c|c|c|c|c|c|c|c}
\toprule
Rounds & 0 &1 &2 &3 &4 &5 &6 &7  \\
\midrule
Provider offer & 17& 16& 15& 14& 14& 14& 13& 13\\
\midrule
Client requirement & 7& 7& 7& 7& 7& 7& 7& 7 \\
\bottomrule
\end{tabular}
\label{tab:intermediate}
\end{table}

\begin{table}[!t]
\centering
\setlength{\tabcolsep}{3pt}
\customTableFont
\caption{Comparison of outputs (final offers) of System 3 and Model 4 for the same inputs (original requirements) from the test dataset. Inputs and outputs are in the format: (number of VCPUs, RAM and Storage).}
\begin{tabular}{c|c|c|c}
\toprule
Index of input data & Input & Output of System 3 & Output of Model 4 \\
\midrule
3 & 7, 56, 320 & 13, 107, 320 &	15, 100, 323\\
\midrule
23 & 11, 85, 230 & 35, 85, 530 & 34, 79, 506\\
\midrule
33 & 13, 47, 720 & 43, 107, 720 & 43, 107, 728  \\
\midrule
53 & 17, 41, 520 & 47, 101, 520 & 48, 100, 525  \\
\midrule
63 & 19, 86, 810 & 49, 86, 810	& 49, 87, 817\\
\bottomrule
\end{tabular}
\label{tab:MLvsNg}
\end{table}

%-------------------------------------------------------------------------------
\subsubsection{Simulate different systems with the same model}
\label{sec:diffsystem}
%-------------------------------------------------------------------------------

We use Model 4 to simulate the five systems in \autoref{tab:systemsettings}. 
In System 5, the $priority$ value of VCPU is set to 1.25, and the other $priority$ values are 1.0. 
The training process of the models and related datasets are consistent with \autoref{sec:diffmodel}.
According to \autoref{tab:systemresults}, Model 4 simulates all systems well. 
Different membership functions, rule sets of different complexity, and cases have very limited impact on the accuracy. 
Furthermore, on Dataset 2, Model 4 takes less than one second to complete all processing. Model 4 increases the processing speed dozens of times while maintaining a similar output to these systems.

We conclude that using the Relu layer in the neural network model can significantly improve the model's ability to simulate fuzzy negotiation systems. 
Neural network models with Relu layers can replace all versions of fuzzy negotiation systems and significantly improve processing speed.

%-------------------------------------------------------------------------------
\subsubsection{Interpretability}
\label{sec:explain}
%-------------------------------------------------------------------------------

%From \autoref{tab:systemresults} and \autoref{fig.comparea}, \autoref{fig.compareb}, \autoref{fig.comparec} in Appendix, we could conclude that the output of Model 4 is very close to that of any negotiation system we use. 
Model 4 simulates all systems well. We use Model 4 to simulate System 3 to further explain interpretability. 
The priority value of VPU is set to 1.5 while the others are 1.0. 
We randomly selected five data points from Dataset 2 as input, and the corresponding output is listed in \autoref{tab:MLvsNg}. 
The outputs of System 3 and Model 4 are so close that we can approximate that they are the same. 
The output of System 3 is interpretable. 
Taking data point 3 as an example, \autoref{tab:intermediate} clearly shows the changes in the number of VCPUs during the seven rounds of negotiation in System 3. 
The same goes for other parameters (RAM and Storage).

Based on the above, it can be concluded that our machine model can obtain good representative results as the fuzzy negotiation system while significantly improving the processing speed. 
The outputs of the fuzzy negotiation system (including the negotiation results of any intermediate rounds) are observable and interpretable. 
Therefore the output of the ML model can also be considered trustworthy and interpretable.

%-------------------------------------------------------------------------------
\section{Related work}
\label{Related work}
%-------------------------------------------------------------------------------

In terms of the application scenarios of auto-negotiation, %there are the following studies recently.
Pierson et al. \cite{pierson2019datazero} explore the application of negotiation in server operations and power control. 
They propose a solution where each side optimizes its objectives, interacting through a negotiation loop to reach a joint agreement and present DATAZERO, a project based on the above idea. 
%They present DATAZERO, a project developing this idea to ensure high availability of IT services, avoiding unnecessary redundancies under the constraints due to the intermittent nature of electrical and cloud services flows. 
Thi et al. \cite{thi2020negotiation} also focus on power management, especially green data centers.
They propose a buyer-supplier negotiation game and algorithm for the problem of optimizing server operations and energy management for data centers entirely powered by renewable energy sources.
%Similar to the proposed concepts, 
Jörn et al. \cite{kunsemoller2014game} also study distributed computing. 
They study the dynamics of on-demand pricing and service usage in a two-stage game model for a monopoly Infrastructure-as-a-Service (IaaS) market and propose a game-theoretic model for the IaaS market.

The key to auto-negotiation implementation is judging whether the offer is accepted. 
A typical way is to use a utility function to measure the offer. 
Scoca et al. \cite{scoca2017smart} define a simple formal language to specify interactions between offers and requests modeled as dSLAC expressions and propose a way of determining the cost of agreeing by relying on utility functions. 
Chokhani et al. \cite{6612196} propose a method that introduces the feature of on-demand resource availability and elasticity using lease auto-negotiation in Haizea. 
Their work adds a new class of lease called Dynamic lease which can be renegotiated according to its current utilization. 
%In the study of control theory, some researchers have recently used fuzzy logic to implement a negotiation system to help control. 
Masero et al. \cite{masero2021hierarchical} propose a fuzzy-based two-layer control architecture. In the lower control layer, there are pairwise negotiations between agents according to the couplings and the communication network.
A coordinator in the upper control layer collects the resulting pairwise control sequences and merges them to compute the final ones. 
Kolomvatsos et al. \cite{kolomvatsos2015adaptive}  propose a fuzzy logic (FL) system with an adaptation technique that updates the FL rule base and membership functions as necessary.
%and is responsible for determining the appropriate actions of the buyer during every negotiation round.

Another key is how to quickly approach the solution of the negotiation or estimate the approximate solution. 
Considering the multi-tenancy market operation mode competing in the cloud environment, Wei et al. \cite{7422131} propose a cloud resource allocation model based on an imperfect information Stackelberg game using a Hidden Markov model (HMM). 
%The HMM is applied to predict the tenants’ bids according to their historical resource demand to save negotiation time. 
Rahman et al. \cite{9097295} propose a negotiation-game-based auto-scaling method where tenants and Network Operator (NO) both engage in the auto-scaling decision via traversal search. They also propose a proactive Machine Learning (ML) based prediction method to predict the tendencies of tenants so that the search range can be reduced to improve efficiency in dynamic scenarios.

Combining the fields of Negotiation, Fuzzy Logic, and Machine Learning is relatively new, especially in distributed computing  infrastructures. Related research is very limited. Carbo et al. \cite{carbo2003machine} study the application of machine learning techniques to compare the privacy of preferences function using fuzzy sets as counter-offers with the classical alternative based on concrete values. They use ML to evaluate the fuzzy counter-offer in one negotiation round. Khuen et al. \cite{khuen2005framework} use ML to update the fuzzy rule base.

In comparison to the above-mentioned related work, our study features the following key points:

\begin{itemize}
\item We have designed and developed a framework for 1-to-1 negotiation to schedule the resource in a distributed infrastructure. Different negotiation cases are supported. 
We apply a fuzzy logic system with novel input parameters and fuzzy rules on the client's agent side to evaluate the offers. 
During the negotiation process, the offers are updated using several new specific algorithms. 
\item We systematically and comprehensively study the feasibility of neural network surrogate models to replace the entire fuzzy negotiation process. 
To the best of our knowledge, this is the only one so far and significantly different from other studies of single-round outcomes.
\end{itemize}

%-------------------------------------------------------------------------------
\section{Conclusion and future direction} 
\label{sec.conclusion}
%-------------------------------------------------------------------------------

We propose an auto-negotiation system for resource scheduling based on fuzzy logic. 
In the system, the client agent uses fuzzy logic to mimic the human response to make decisions, while the provider agent dynamically updates the offer according to the agreed terms.
We compare the impact of different member functions, fuzzy rule sets, and negotiation scenario cases on the final offers.
The framework also includes neural network models to fit the fuzzy negotiation systems above.
The trained model provides a much faster response time for requests than the fuzzy negotiation system, which is very important in the real-world cloud environment.

Based on the presented results, we can conclude that the proposed system can utilize resources more efficiently. It is both interpretable and efficient and has practical value in business activities. 

In the future, we would like to extend our work toward the adaptive fuzzy rules. The fuzzy rule set depends largely on prior knowledge in the fuzzy negotiation system. 
Fuzzy rule sets with adaptive capabilities can modify, delete, or add new fuzzy rules during negotiation, compensating for the inaccuracy of prior knowledge and preset rule sets. 
Together with Fuzzy rules, we would also like to develop new architectures for machine learning models. 
We would like to include making the model aware of handling client priorities, which can significantly improve the negotiation process.  

Machine learning models also remain flawed. For the current machine learning model, the input features we use are still too few. 
When the client modifies the parameters (such as $priority$), we need to retrain and deploy the model. 
In the future, we consider adding parameters such as $priority$ to the input features and increasing the number of input features. 
Models trained with more features are more general and easier to transfer and deploy.

%-------------------------------------------------------------------------------
\section*{Acknowledgment}
%-------------------------------------------------------------------------------

We want to acknowledge the Swedish National Infrastructure for Computing (SNIC) for providing cloud resources and support from the eSSENCE strategic collaboration in e-science.

%-------------------------------------------------------------------------------
\begin{small}
\bibliographystyle{plain}
\bibliography{sample}    
\end{small}
%-------------------------------------------------------------------------------

%-------------------------------------------------------------------------------
\appendix
\section*{Appendix}
%-------------------------------------------------------------------------------

\begin{algorithm}[!ht]
\begin{footnotesize}
  \setcounter{algorithm}{0}
  \caption{\small Generating the quantity of one parameter in the first offer (Provider agent)}
  \label{alg:1}
   \algorithmicrequire\\
    \hspace*{0.02in} $n_0$: the quantity in the original requirement;\\
    \hspace*{0.02in} $T_{1-2}$: the threshold value of the quantity of the level-1 price and the level-2 price;\\
    \hspace*{0.02in} $T_{2-3}$: the threshold value of the quantity of the level-2 price and the level-3 price;\\  
   \algorithmicensure\\  
    \hspace*{0.02in} $n_1$: the quantity in the first offer;\\
  \begin{algorithmic}[1]   
       \If {$0<n_0<=T_{1-2}$}
       \State {$n_1=n_0+T_{1-2}$}
       \ElsIf {$T_{1-2}<n_0<=T_{2-3}$}
       \State {$n_1=n_0+T_{2-3}$}
       \Else
       \State {$n_1=n_0$}
       \EndIf
  \end{algorithmic}
\end{footnotesize}
\end{algorithm}

\begin{algorithm}[!ht]
\begin{small}
  \caption{\small Generating the advice (Client agent)} % 名称
  \label{alg:2}
   \algorithmicrequire\\
    \hspace*{0.02in} $n_0^{VCPU},n_0^{RAM},n_0^{Storage}$: the quantities of VCPU, RAM, storage in the original requirement;\\
    \hspace*{0.02in} $n_c^{VCPU},n_c^{RAM},n_c^{Storage}$: the quantities of VCPU, RAM, storage in the current offer;\\
    \hspace*{0.02in} $d_{max}$: the maximum deviation value;\\
   \algorithmicensure\\  
    \hspace*{0.02in} $Advice$: the parameter the client agent advises to reduce or \emph{random};\\
  \begin{algorithmic}[1]
       \State {Compute the ratios:}
       \Statex {$r_{c/0}^{VCPU}=n_c^{VCPU}/n_0^{VCPU}, r_{c/0}^{RAM}=n_c^{RAM}/n_0^{RAM},$}
       \Statex {$r_{c/0}^{Storage}=n_c^{Storage}/n_0^{Storage}$}
       \State {Find the minimum ratio:} 
       \Statex {$r_{min} = min(r_{c/0}^{VCPU},r_{c/0}^{RAM},r_{c/0}^{Storage})$}
       \State {Find the maximum ratio:} 
       \Statex {$r_{max} = max(r_{c/0}^{VCPU},r_{c/0}^{RAM},r_{c/0}^{Storage})$}
       \State {Compute the deviation: $d=r_{max}/r_{min}$}
       \If {$d<d_{max}$}
       \State {$Advice = 'random'$}
       \Else
       \State {Find the parameter corresponding to the maximum ratio: $P_{max}$}
       \State {$Advice = P_{max}$}
       \EndIf
  \end{algorithmic}
\end{small}
\end{algorithm}

\begin{algorithm}[!ht]
\begin{footnotesize}
  \renewcommand{\thealgorithm}{3.a}
  \caption{\small Provider agent's algorithm for selecting the parameter to reduce (Case a)}
  \label{alg:3a}
   \algorithmicrequire\\
    \hspace*{0.02in} $n_0^{VCPU},n_0^{RAM},n_0^{Storage}$: the quantities of VCPU, RAM, storage in the original requirement;\\
    \hspace*{0.02in} $c^{VCPU},c^{RAM},c^{Storage}$: the amount of change per round;\\
   \algorithmicensure\\  
    \hspace*{0.02in} $P_{reduce}$: the parameter the provider agent selects to reduce;\\
  \begin{algorithmic}[1]   
   \State {Compute the number of rounds each parameter needs:}
   \Statex {$rd^{VCPU}=n_0^{VCPU}/c^{VCPU}, rd^{RAM}=n_0^{RAM}/c^{RAM},$}
   \Statex {$rd^{Storage}=n_0^{Storage}/c^{Storage}$}
   \State {Compute the total number of rounds:}
   \Statex {$rd^{total}=rd^{VCPU}+rd^{RAM}+rd^{Storage}$}
   \State {Compute the probability of each parameter:}
   \Statex {$pb^{VCPU}=rd^{VCPU}/rd^{total},  pb^{RAM}=rd^{RAM}/rd^{total},$}
   \Statex {$pb^{Storage}=rd^{Storage}/rd^{total}$}
   \State {According to the above probability, randomly select a parameter to reduce: $P_{reduce}$}
  \end{algorithmic}
\end{footnotesize}
\end{algorithm}

\begin{algorithm}[!ht]
\begin{footnotesize}
  \renewcommand{\thealgorithm}{3.b}
  \caption{\small Provider agent's algorithm for selecting the parameter to reduce (Case b)}
  \label{alg:3b}
   \algorithmicrequire\\
    \hspace*{0.02in} $n_0^{VCPU},n_0^{RAM},n_0^{Storage}$: the quantities of VCPU, RAM, storage in the original requirement;\\
    \hspace*{0.02in} $c^{VCPU},c^{RAM},c^{Storage}$: the amount of change per round;\\
    \hspace*{0.02in} $Advice$: the advice sent from the client agent;\\
   \algorithmicensure\\  
    \hspace*{0.02in} $P_{reduce}$: the parameter the provider agent selects to reduce;\\
  \begin{algorithmic}[1]
   \If {$Advice = 'random'$}
   \State {Jump to \autoref{alg:3a}}
   \Else
   \State {$P_{reduce} = Advice$}
   \EndIf
  \end{algorithmic}
\end{footnotesize}
\end{algorithm}

\begin{algorithm}[!ht]
\begin{footnotesize}
  \renewcommand{\thealgorithm}{3.c}
  \caption{\small Provider agent's algorithm for selecting the parameter to reduce (Case c)}
  \label{alg:3c}
   \algorithmicrequire\\
    \hspace*{0.02in} $n_0^{VCPU},n_0^{RAM},n_0^{Storage}$: the quantities of VCPU, RAM, storage in the original requirement;\\
    \hspace*{0.02in} $c^{VCPU},c^{RAM},c^{Storage}$: the amount of change per round;\\
    \hspace*{0.02in} $Advice$: the advice sent from the client agent;\\
    \hspace*{0.02in} $pr^{VCPU},pr^{RAM},pr^{Storage}$: the priority of each parameter;\\
   \algorithmicensure\\  
    \hspace*{0.02in} $P_{reduce}$: the parameter the provider agent selects to reduce;\\
  \begin{algorithmic}[1]
   \If {$Advice = 'random'$}
   \State {Compute the number of rounds each parameter needs:}
   \Statex {\hspace*{0.22in}$rd_0^{VCPU}=n_0^{VCPU}/c^{VCPU},$}  
   \Statex {\hspace*{0.22in}$rd_0^{RAM}=n_0^{RAM}/c^{RAM},$}
   \Statex {\hspace*{0.22in}$rd_0^{Storage}=n_0^{Storage}/c^{Storage}$}
   \State {Compute the relative adjusted number of rounds:}
   \Statex {\hspace*{0.22in}$rd^{VCPU}=rd_0^{VCPU}/pr^{VCPU},$}
   \Statex {\hspace*{0.22in}$rd^{RAM}=rd_0^{RAM}/pr^{RAM},$}
   \Statex {\hspace*{0.22in}$rd^{Storage}=rd_0^{Storage}/pr^{Storage}$}
   \State {Compute the total number of rounds: }
   \Statex {\hspace*{0.22in}$rd^{total}=rd^{VCPU}+rd^{RAM}+rd^{Storage}$}
   \State {Compute the probability of each parameter:}
   \Statex {\hspace*{0.22in}$pb^{VCPU}=rd^{VCPU}/rd^{total},  pb^{RAM}=rd^{RAM}/rd^{total},$}
   \Statex {\hspace*{0.22in}$pb^{Storage}=rd^{Storage}/rd^{total}$}
   \State {Randomly select $P_{reduce}$ according to the above probability }
   \Else
   \State {$P_{reduce} = Advice$}
   \EndIf
  \end{algorithmic}
\end{footnotesize}
\end{algorithm}

\end{document}